\documentclass[twocolumn,showpacs,preprintnumbers,amsmath,amssymb,aps]{revtex4}

\usepackage{graphicx}
\usepackage{dcolumn}
\usepackage{bm}

\begin{document}

\title{Multiple Choice Minority Game}

\author{F. K. Chow and H. F. Chau}
\affiliation{
  Department of Physics, University of Hong Kong, Pokfulam Road, Hong Kong
}
\date{\today}

\begin{abstract}
Minority game is a model of heterogeneous players who think inductively. In 
this game, each player chooses one out of two alternatives every turn and those 
who end up in the minority side wins. It is instructive to extend the minority 
game by allowing players to choose one out of many alternatives. Nevertheless, 
such an extension is not straight-forward due to the difficulties in finding a 
set of reasonable, unbiased and computationally feasible strategies. Here, we 
propose a variation of the minority game where every player has more than two 
options. Results of numerical simulations agree with the expectation that our 
multiple choices minority game exhibits similar behavior as the original 
two-choice minority game.
\end{abstract}

\maketitle

\section{\label{Intro}Introduction}
Complex adaptive systems have drawn attention among statistical physicists in 
recent years. The study of such systems not only provides invaluable insight 
into the non-trivial global behavior of a population of competing agents, but 
also has potential application in economics, biology and finance. Moreover, we 
can study complex adaptive systems from the perspective of statistical physics.

The El Farol bar problem \cite{Min1}, which was proposed by W. B. Arthur in 
1994, has greatly influenced and stimulated the study of complex adaptive 
systems in the last few years. It describes a system of $N$ agents deciding 
independently in each week whether to go to a bar or not on a certain night. 
As space is limited, the bar is enjoyable only if it is not too crowded. 
Agents are not allowed to communicate directly with each other and their 
choices are not affected by previous decisions. The only public information 
available to agents is how many agents came in last week. To enjoy an 
uncrowded bar, each agent has to employ some hypotheses or mental models to 
guess whether one should go or not. With bounded rationality, all agents use 
inductive reasoning rather than perfect, deductive reasoning since the system 
is ill-defined. In other words, agents act upon their currently most credible 
hypothesis based on the past performance of their hypotheses. Consequently, 
agents can interact indirectly with each other through their hypotheses in use. 
The emerging system is thus both evolutionary and complex adaptive in the 
presence of those inductive reasoning agents.   

Inspired by the El Farol bar problem, Challet and Zhang put forward the 
Minority Game (MG) \cite{Min2,Min3}. It is a toy model of $N$ inductive 
reasoning players who have to choose one out of two alternatives independently 
at each time step. Those who end up in the minority side (that is, the choice 
with the least number of players) win. The only public information available to 
all players in the MG is the winning alternatives of the last $M$ passes, known 
as the history. Players have to employ some strategies based on the history to 
guess the winning choice. In fact, they may employ more than one strategy 
throughout the game. More precisely, every player picks once and for all $S$ 
randomly drawn strategies from a suitably chosen strategy space before playing 
the game. The performance of each strategy will be recorded during the game. 
Then players make decision according to their current best performing strategy 
at every time step.

In spite of its simplicity, MG displays a remarkably rich emergent collective 
behavior. Numerical simulations showed that there is a second order phase 
transition between a symmetric phase and an asymmetric phase 
\cite{Min4,Min5,Min6}. There is no predictive information about the next 
minority group available to agent's strategies in the symmetric phase, whereas 
there is predictive information available to agent's strategies in the 
asymmetric phase. MG addresses the interaction between agents and public 
information, that is, how agents react to public information and how the 
feedback modify the public information itself. Later works revealed that the 
dynamics of the system in fact minimizes a global function related to market 
predictability \cite{Min7,Min8}. Therefore, the MG can be described as a 
disordered spin system \cite{Min7,Min8}. Hart and his coworkers found that the 
fluctuations arising in the MG is controlled by the interplay between crowds of 
like-minded agents and their anti-correlated partners \cite{Min9,Min10}. Since 
MG is a prototype to study detailed pattern of fluctuations, it plays a 
dominant role in economic activities like the market mechanism. In order to 
learn more on MG-like systems, much effort was put on the extension of the MG 
model, such as the introduction of evolution \cite{Min11} and the modification 
of MG for modelling the real market \cite{Min12,Min13}.

In the real world, however, an agent usually has more than two options. For 
example, people decide where to dine or which share to buy from a stock market. 
Consequently, it is worthwhile to investigate the situation where players have 
more than two choices especially when the number of choices is large. Indeed, 
D'Hulst and Rodgers \cite{Min14} has written a paper on the three-choice 
minority game. In their paper, a symmetric and an asymmetric three-sided models 
are introduced. Their symmetric model is only a model which mimics the cyclic 
trading between three players using the same strategy formalism as MG and thus 
cannot be extended to realistic cases with a large number of choices. 
Furthermore, their asymmetric model is nothing but the original minority game 
with the possibility allowing players not to participate in a turn. Hence, the 
player's choice is not symmetric.

Recently, Ein-Dor \emph{et al.} proposed a multichoice minority game based on
neural network \cite{Min15}.
Their model generalizes the El Farol bar problem to the case of
multiple choice. Nevertheless, it differs quite significantly from the MG of
Challet and Zhang as each player has only one strategy to use and that strategy
evolves according to its performance. Besides, no phase transition of any kind 
is observed in Ein-Dor \emph{et al.}'s model. 
Chau and Chow also proposed another multichoice minority game model based on
MG \cite{Min16}. In this model, players choosing their strategies from
a reduced strategy space consists of anti-correlated and uncorrelated
strategies only.

In this paper, we propose a new model called the multiple choices minority game 
(MCMG) with a neural network flavor. 
It is a variation of the MG where all heterogeneous, inductive thinking 
players have more than two choices. Just like the original MG, strategies in MCMG 
are not evolving and they are picked in each turn according to their current 
performance. In Section \ref{SecModel}, the MCMG model 
are explained in detail. Results of numerical simulation of our model are 
presented and discussed in Section \ref{SecResults}. We also compare our 
results with those of the original MG as well. In the Section \ref{SecConc}, we 
deliver a brief summary and an outlook of our work.

\section{The Model}
\label{SecModel}
Let us consider a repeated game of a population of $N$ players. At each time 
step, every players has to choose one of $N_c$ rooms/choices 
{\it independently}. Here, we assume that $N_c$ is a prime power and we
identify the $N_c$ rooms with elements in the finite field $GF(N_c)$.
We represent the choice of the $i$th player at time $t$ by $\chi_i(t)$ which
only ``takes on'' the $N_c$ different rooms.  
Those players in the room with the least number of players, that is, in the 
{\it minority} side, win. The winning room at time $t$ is the publicly known 
output of the game $\Omega(t)$. The players of the winning room will gain one 
unit of wealth while all the others lose one. So the wealth of the $i$th
player, $w_i(t)$, is updated by
\begin{equation}
  w_i(t+1) = w_i(t) + 2\delta(\chi_i(t) - \Omega(t)) - 1 ,
\end{equation}
where $\delta (0) = 1$ and $\delta (x) = 0$ whenever $x\neq 0$.
Note that the output of the last $M$ steps, $\mu(t) \equiv ( \Omega(t-M+1), 
\ldots, \Omega(t-1) )$, is the {\it only} public information available to all 
players. Therefore, players can only interact indirectly with each other 
through the history $\mu(t)$ which can take on $N_c^M$ different values. 

Aimed at maximizing one's own wealth, each player has to employ some strategies 
to predict the trend of the output of the game. But how to define a strategy? 
For a minority game with more than two choices, it is not effective to 
formulate the strategy as in MG. Recall that in MG, a strategy is defined as a 
number of set of choices corresponding to different histories. In other words, 
a strategy is a map sending each $\mu(t)$ to the choice $0/1$. Therefore, there 
are $2^{2^M}$ different strategies in the full strategy space for MG. 

Similar numerical results of the fluctuation arising in the MG are obtained if 
strategies are drawn from the reduced strategy space instead of the full 
strategy space \cite{Min2,Min4}. Hence, the reduced strategy space plays a 
fundamental role in the properties of the fluctuation arising in the MG. The 
reduced stratgey space is formed by strategies which are significantly different 
from each other. Given a strategy $s$, only its uncorrelated and 
anti-correlated strategies are significantly different from it.
Indeed, the reduced strategy space is composed of two ensembles of mutually 
uncorrelated strategies where the anti-correlated 
strategy of any strategy in one ensemble always exists in the other ensemble. 
For MG, there are $2^{M+1}$ different strategies in the reduced strategy space 
\cite{Min2,Min4}. 

For a minority game with $N_c$ rooms using the form of strategy in MG, there 
are $N_c^{N_c^M}$ different strategies in the full strategy space while there 
are $N_c^{M+1}$ different strategies in the reduced strategy space provided 
that $N_c$ is a prime power. 
Because the 
strategy space size increases rapidly as $N_c$ increases, strategies will 
quickly get out of control for large $N_c$. Consequently, we would like to 
define the strategy in a different way. 

In our game, we do not restrict players to have only ``good strategies'' since 
each player (who thinks inductively) does not know whether a strategy is good 
or not before commences the game. ``Good strategy'' is time dependent. In fact, 
players will adapt with each other in order to use those ``good strategies''. 
Therefore, all strategies must be {\it uniform} in the following sense:
\newcounter{c}
\begin{list}
  {\arabic{c})}{\usecounter{c}}
\item any input $\mu(t)$ can produce any output $\chi_i(t)$,
\item there are same probability for any output produced by any input.
\end{list}
Indeed, it is not completely clear that whether those strategies used in the
Ein-Dor \emph{et al.}'s model are uniform in the above sense.
With the above consideration, a strategy $s$ consists of weights
($\omega^s_1, \ldots, \omega^s_M$) $\in GF(N_c)^M$ and a uniform random
variable called bias $\rho_s \in GF(N_c)$ satisfying the following
condition:
\begin{equation}
  \sum_{j=1}^{M}\omega^s_j = \eta
\end{equation} 
where $\eta$ is a fixed constant in $GF(N_c)$.
Note that all the arithemetic used to generate a strategy and the corresponding
choice (including Eq.~(2) above and Eq.~(3) below) are performed in the finite 
field $GF(N_c)$. 
The choice is defined to be the sum of the weighted sum of the last $M$ output
plus the bias. Namely, the choice of the $i$th player using the strategy $s$ with
history $\mu(t)$ is given by: 
\begin{equation}
  \chi^{\mu(t)}_{i,s_i(t)} = \rho_s + \sum_{j=1}^{M} \omega^s_j \Omega(t-j)
\end{equation}
Physically, the weight $\omega^s_j$ represents the importance of the output of 
the game of $j$th pass before on the choice. It is obvious that the strategies
of MCMG all fulfil the uniform criteria which was mentioned before. 

Since the weights of a strategy are $M$ independent variable in $GF(N_c)$
which satisfy a single constraint, namely Eq.~(2), there are $N_c^{M-1}$
different combinations of weights $\omega_j^s$. 
Moreover, it can be shown that strategies with the same set of weights are 
anti-correlated with each other while the others are uncorrelated with each
other (see Ref.~\cite{Min16}).
Hence, it follows that both the full and reduced strategy space size in MCMG
is equal to $N_c^M$. 
Thus, the strategy space size of MCMG is much smaller than that of MG. 

In our model, every player picks and sticks to $S$ randomly drawn strategies 
before commences the game. We denote the strategies of the $i$th player by 
$(s^{(i)}_1, \ldots, s^{(i)}_S)$. But how does a player decide which strategy 
is the best? Players use the virtual score, which is just the hypothetical 
profit for using a single strategy in the game, to estimate the performance of 
a strategy. In each pass, the virtual score of a strategy $s$ of the $i$th player, 
$U^i_s(t)$, is updated by
\begin{equation}
  U^i_s(t) = U^i_s(t-1) + 2\delta(\chi^{\mu(t)}_{i,s} - \Omega(t)) - 1
\end{equation}
where $\chi^{\mu(t)}_{i,s}$ is the choice of the $i$th player under history 
$\mu(t)$ using the strategy $s$. Each player uses one's own strategy with the 
highest virtual score. 
Although our MCMG model is quite similar to the MG model, there are two main 
differences, namely, in the number of choices of players and in the formalism of 
strategies.  

In our game, the aim of each player is to maximize one's own wealth which can 
be in turn achieved by the maximization of the global profit. So the quantity 
of interest is 
\begin{eqnarray}
  \sigma_j^{2} &=& \langle (A_j(t))^2 \rangle - \langle A_j(t) \rangle^2
  \nonumber \\
  &\equiv& \langle (A_j(t) - N/N_c)^2 \rangle , 
\end{eqnarray}
namely, the variance of the attendance of room $j$, where the attendance of 
a room is just the number of people chosen that room. Indeed, the maximum 
global profit will be achieved if and only if the largest possible minority 
size $\lfloor N/N_c \rfloor$ is attained. So the expected attendance of all 
rooms should be equal to $N/N_c$ for players to gain as much as possible. 
Accordingly, the variance of the attendance of a room represents the loss of 
players in the game.

In order to investigate the significance of the strategies, we would like to 
compare the variance with the coin-toss case value in which all the players 
make their decision simply by tossing an unbiased coin. It is easy to check 
that the probability for the attendance $A_j(t)$ equal to $x$ in the coin-toss 
case is given by
\begin{equation}
  p(A_j(t)=x) = \frac{{}^N\!C_x (N_c-1)^{N-x}}{\sum_{x=1}^{N} 
  {}^N\!C_x (N_c-1)^{N-x}} ,
\end{equation}
where ${}^N\!C_x = N!/[x!(N-x)!]$. So the expectation of $A_j$ and ${A_j}^2$ in 
the coin-toss case are given by 
\begin{equation}
  \langle A_j \rangle = \frac{\sum_{x=1}^{N}x \ {}^N\!C_x (N_c-1)^{N-x}}
  {\sum_{x=1}^{N} {}^N\!C_x (N_c-1)^{N-x}} 
\end{equation}
and
\begin{equation}
  \langle {A_j}^2 \rangle  = \frac{\sum_{x=1}^{N}x^2 \ {}^N\!C_x (N_c-1)^{N-x}}
  {\sum_{x=1}^{N} {}^N\!C_x (N_c-1)^{N-x}} .
\end{equation} 
   
\section{Results of numerical simulations}
\label{SecResults}
In all the simulations, each set of data was 
taken for 1,000 independent runs. In each run, we took the average values on 
15,000 steps after running 10,000 steps for equilibrium starting from 
initialization. 

\subsection{Comparison of two-room MCMG with MG}

We first want to investigate if the performance of players are different in MG 
and the two-room MCMG. We applied the two models to study the properties of the 
mean attendance as a function of the control parameter $\alpha$ as shown in
figure~1. 
The control parameter $\alpha$ measures the ratio of the reduced strategy space
size to the number of strategies at play. Specifically, $\alpha = 2^{M+1}/NS$
for the MG and $\alpha = N_c^M/NS$ for the MCMG.
We have only studied the properties of the attendance of one 
of the rooms as the attendance of the two rooms have the same behavior due to 
symmetry \cite{Min1}.

In both MG and the two room MCMG, the mean attendance always fluctuates around 
the expected value $N/2$ no matter how large is the control parameter (see 
figure~1). Therefore, we believe that players have the same performance in 
both MG and the two room MCMG if we only consider the mean attendance.
To further investigate the difference of the performance of players in MG and 
the two-room MCMG, we studied the variance of the attendance as a function of 
the control parameter which was shown in figure~2.

Before making comparsion with that in the MCMG, let us first take a brief 
review on the properties of the variance in MG as shown in figure~2 
\cite{Min4,Min5,Min6}. In MG, there is a great variance $\sigma^{2}$ when the 
control parameter $\alpha$ is small. It is because players use very similar 
strategies when the reduced strategy space size ($ = 2^{M+1}$) is much smaller 
than the number of strategies at play ($ = NS$). Such overcrowding of strategies 
will lead to small minority size and also great fluctuation of the attendance. 
When the reduced strategy space size increases, the variance decreases rapidly 
since players are able to cooperate more with less overcrowding effect. 
Subsequently, the variance attains a minimum when the number of strategies at 
play is approximately equal to the reduced strategy space size. In fact, 
maximum cooperation is attained when all the strategies in the reduced strategy
space are used by players. As the slope of 
the variance changes discontinuously at the minimum point, it strongly suggests 
that there is a second order phase transition. The phase transition was also 
confirmed by analytical method \cite{Min7,Min17}. After the minimum point, the 
variance increases and gradually tends to the coin-toss case value when the 
control parameter $\alpha$ increases further. It is due to more and more 
insufficient sampling of the reduced strategy space when the reduced strategy 
space size become much and much larger than the number of strategies at play. 
Consequently, there is less and less difference between choosing by random or 
using strategies. In addition, the properties of the variance are found to be 
similar for different number of strategies $S$. 

\begin{figure}[!t]
\includegraphics[scale = 0.28, bb=520 50 300 600]{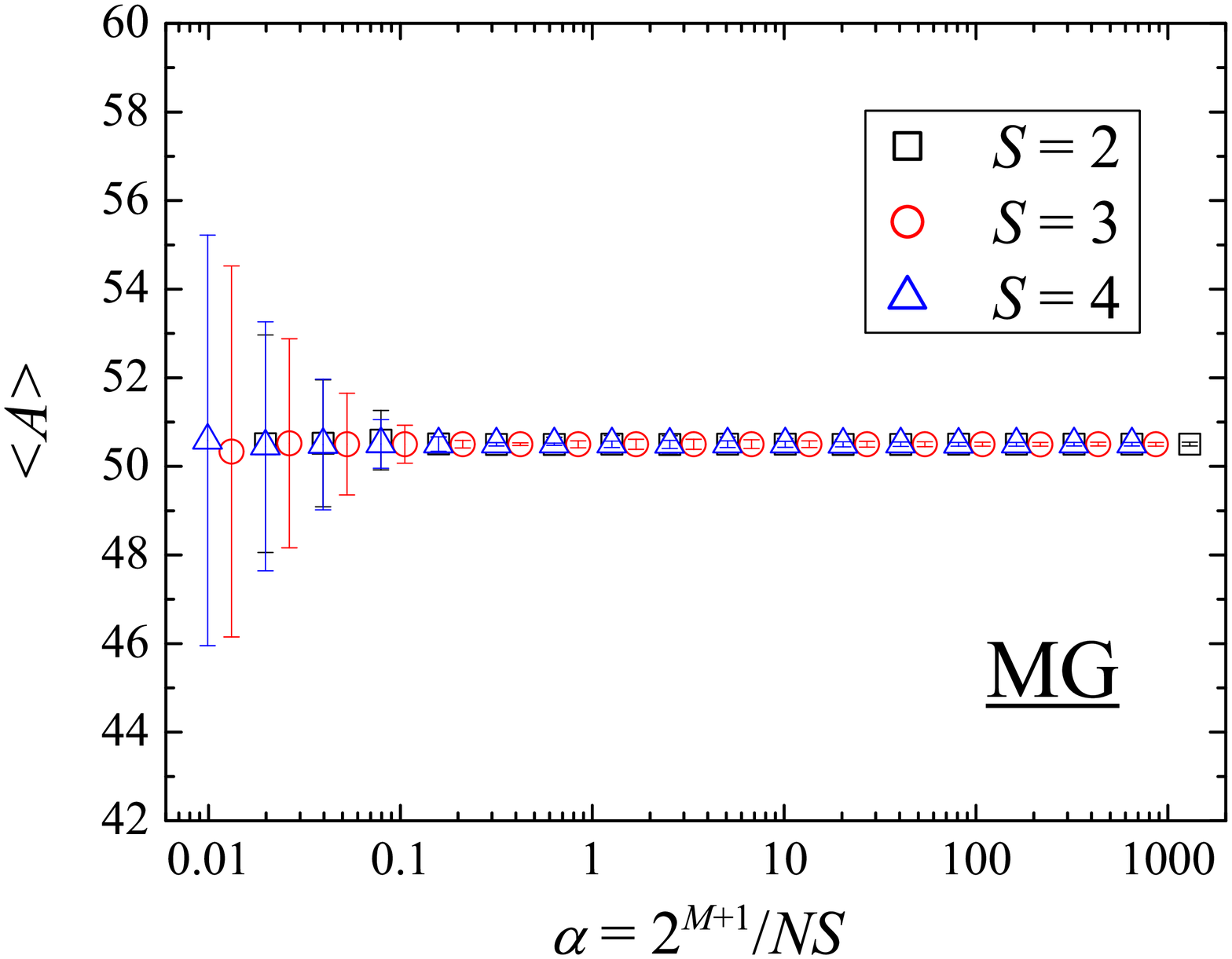}
\end{figure}
\begin{figure}[!t]
\vspace*{-4mm}
\includegraphics[scale = 0.28, bb=520 50 300 600]{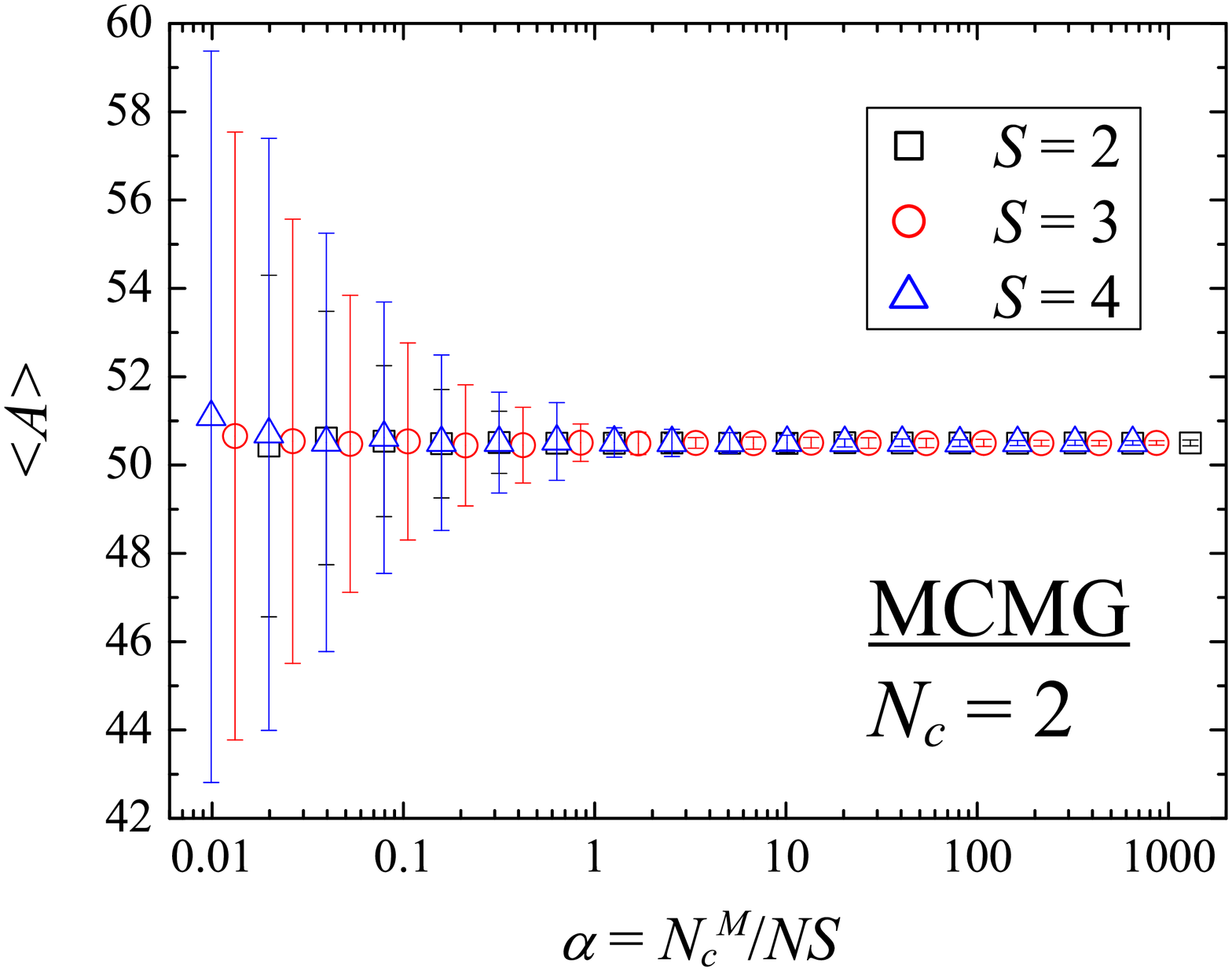}
\caption[1]{The mean attendance $\langle A \rangle$ versus the control parameter 
$\alpha$ in MG and MCMG with $N_c = 2$ where $N = 101$.}
\end{figure}

\begin{figure}[!t]
\includegraphics[scale = 0.28, bb=520 50 300 600]{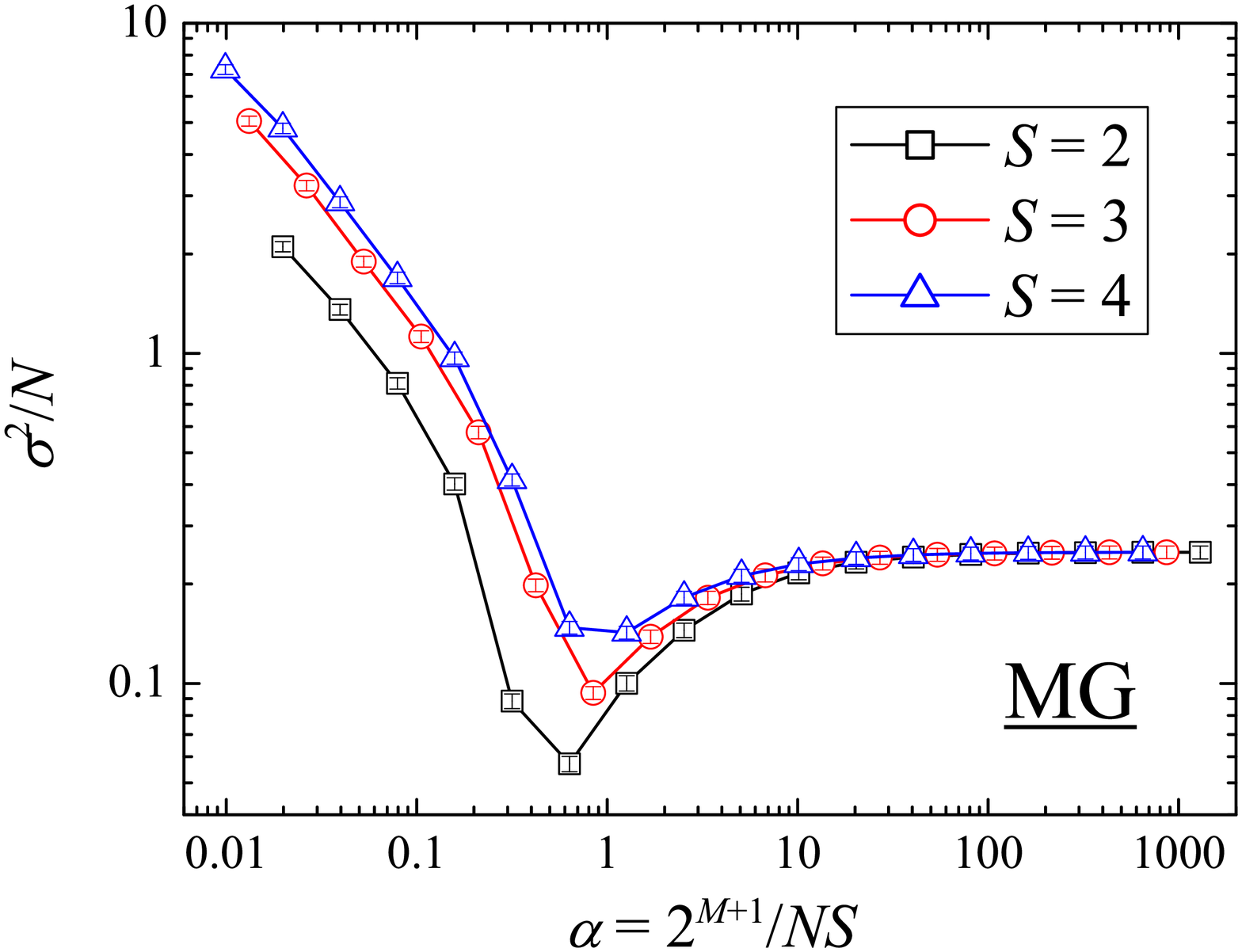}
\end{figure}
\begin{figure}[!t]
\vspace*{-7mm}
\includegraphics[scale = 0.28, bb=520 50 300 600]{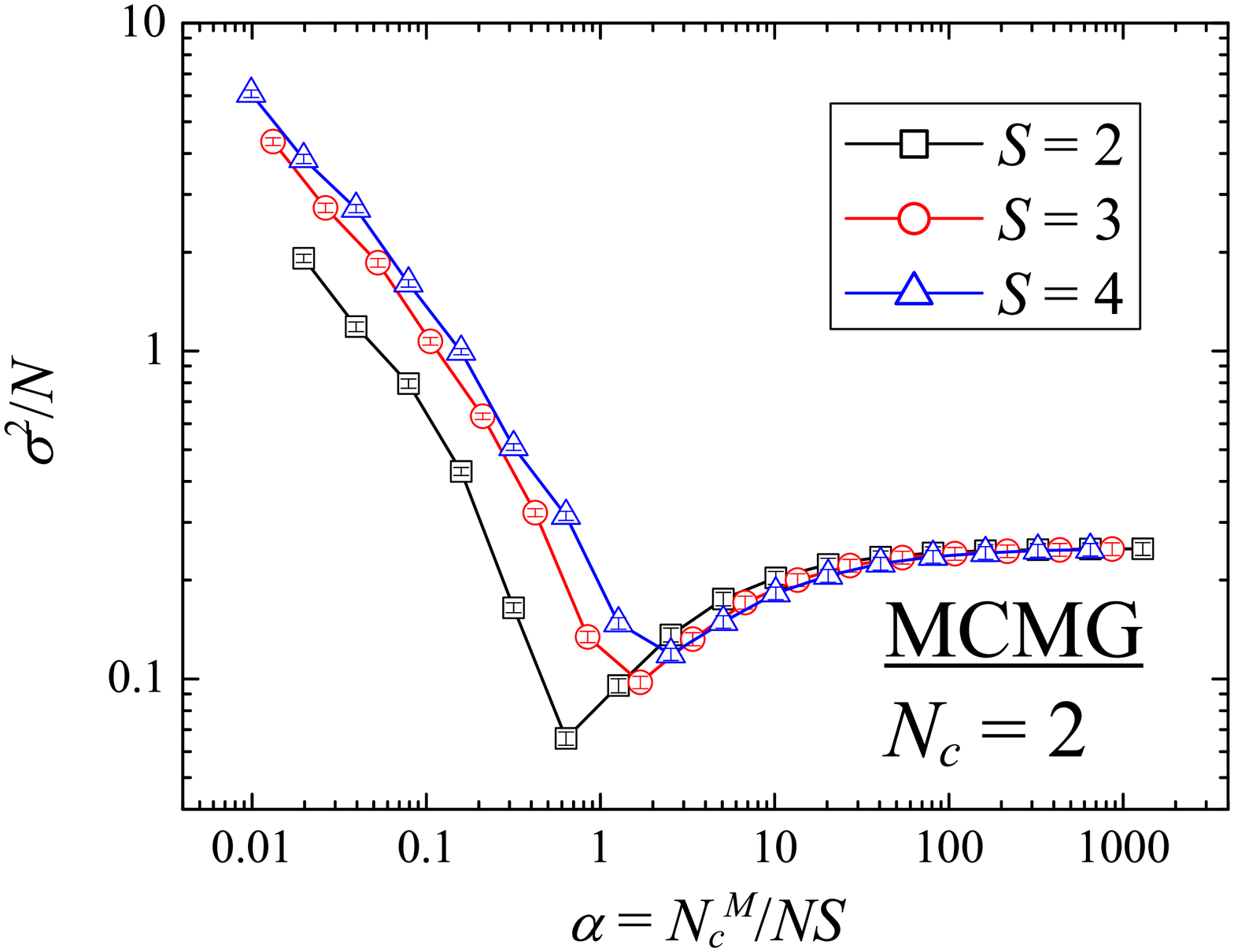}
\caption[1]{The variance $\sigma^2$ versus the control parameter $\alpha$ in MG 
and MCMG with $N_c = 2$ where $N = 101$.}
\end{figure}

For the MCMG, the variance of the attendance of a room, $\sigma^{2}$, exhibits 
similar behavior as a function of the control parameter $\alpha$ ($ = N_c^M/NS$
for MCMG) to that in MG no matter what the number of strategies $S$
is. For example, the variance tends to the coin-toss case value for sufficiently 
large control parameter. Moreover, there is again an indication of 
a second order phase transition. To check if the phase transition is second 
order or not, we calculate the order parameter \cite{Min7,Min16}
\begin{equation}
\theta = \frac{1}{N_c^M} \sum_\mu{ \left\{ \sum_\Omega{ \left[ \langle p(\Omega|\mu) 
\rangle - \frac{1}{N_c} \right]^2 } \right\} } 
\end{equation}
where $\langle p(\Omega|\mu) \rangle$ denotes the conditional time average of 
the probability for $\Omega(t) = \Omega$ given that $\mu(t) = \mu$. In fact, 
the order parameter measures the bias of player's decision to any choice for 
individual history.

\begin{figure}[!h]
\includegraphics[scale = 0.28, bb=520 50 300 600]{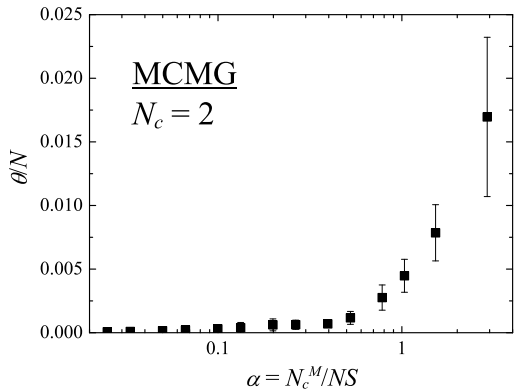}
\caption[1]{The order parameter $\theta$ versus the control parameter $\alpha$ 
in MCMG with $N_c = 2$ where $M = 6$.}
\end{figure}

Figure~3 shows that the order parameter vanishes when the 
control parameter is smaller than its value corresponding to minimum 
variance. As a result, we confirm that the phase transition is a second order 
one. Although our two-room MCMG model does not exactly coincide with the MG 
model, the variance has almost the same properties as a function of the 
control parameter in both models. However, the behaviour of the variance as a 
function of the memory size $M$ are different in MG and MCMG because the 
reduced strategy space size are different in the two models. 

\subsection{The attendance of different rooms in MCMG}

\begin{figure}[!h]
\vspace*{-2mm}
\includegraphics[scale = 0.28, bb=420 50 300 600]{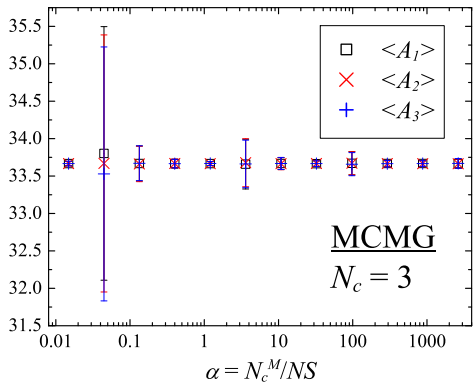}
\caption[1]{The mean attendance $\langle A_j \rangle$ of different rooms versus 
the control parameter $\alpha$ in MCMG with $N_c = 3$ where $S = 2$ and $N = 101$.}
\end{figure}
\begin{figure}[!h]  
\vspace*{-6mm}
\includegraphics[scale = 0.28, bb=420 50 300 600]{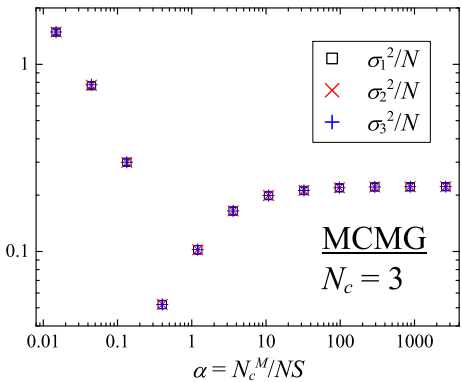}
\caption[1]{The variance $\sigma_j^2$ of different rooms versus the control 
parameter $\alpha$ in MCMG with $N_c = 3$ here $S = 2$ and $N = 101$.}
\end{figure}

Here, we want to study if the properties of the attendance of different rooms 
are different or not in the MCMG. Figures~4 and~5 show the mean attendance and 
the variance of the attendance of different rooms versus the control parameter 
$\alpha$. 
 
We found that the behavior of the attendance of different rooms are almost the 
same in MCMG with $N_c \ge 2$ because there is same probability for any 
choice to be the minority side. As we only want to focus on the study of the 
attendance of a room, not on their high order correlation, so we will stick to 
one of them from now on.

\subsection{The attendance in MCMG with different $N_c$}

Now, we investigate the properties of the attendance for MCMG with different 
number of rooms $N_c$. Figure~6 depicts the dependence of the mean attendance 
on the control parameter $\alpha$ for MCMG with $N_c = 3$ to $7$. In MCMG 
with different $N_c$, the mean attendance fluctuates around $N/N_c$ 
irrespective of the control parameter $\alpha$. It is resonable as the 
maximum global profit will be achieved if and only if the largest possible 
minority size $\lfloor N/N_c \rfloor$ is attained.
 
\begin{figure}[h]
\includegraphics[scale = 0.28, bb=520 50 300 600]{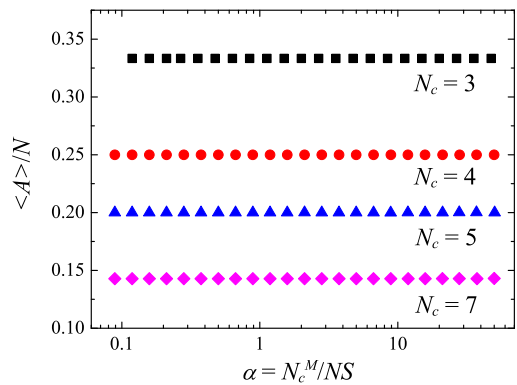}
\caption[1]{The mean attendance $\langle A \rangle$ versus the control 
parameter $\alpha$ in MCMG with different $N_c$ where $S = 2$.}
\end{figure}

We also studied the dependence of the variance of the attendance on the control 
parameter $\alpha$ for MCMG with $N_c = 3$ to $7$ as shown in figure~7. In 
this figure, we have divided the variance by $N/N_c$ ($\approx$ largest possible 
minority size) in order to have an objective comparison of the variance for 
different $N_c$.

\begin{figure}[h]
\includegraphics[scale = 0.28, bb=520 50 300 600]{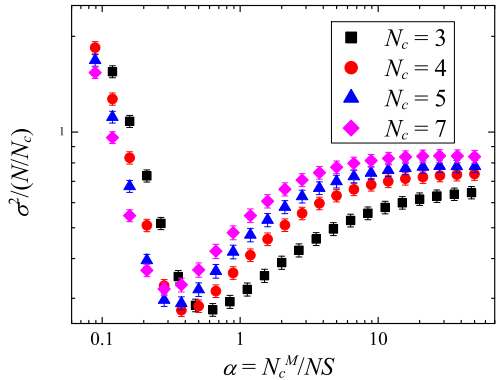}
\caption[1]{The variance $\sigma^2$ versus the control parameter $\alpha$ in 
MCMG with different $N_c$ where $S = 2$.}
\end{figure}

No matter what is the value of $N_c$, there is always a cusp of the variance 
which strongly suggests the occurence of a second order phase transition (see 
figure~7). We calculate the order parameter $\theta$ (introduced in Eq.(10)) to 
identify the order of the phase transition as shown in figure~8. In MCMG with 
different $N_c$, the order parameter vanishes when the control parameter
is smaller than its value at the cusp. Therefore, we conclude that there is a 
second order phase transition in MCMG with different $N_c$.

\begin{figure}[h]
\includegraphics[scale = 0.28, bb=520 50 300 600]{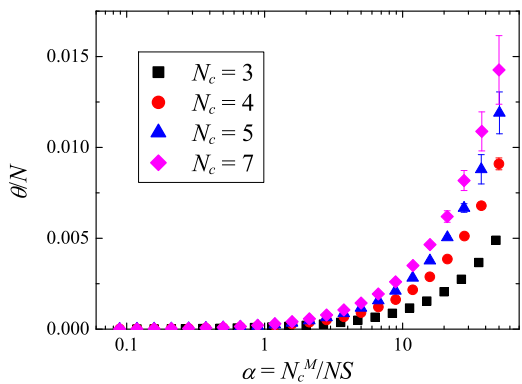}
\caption[1]{The order parameter $\theta$ versus the control parameter $\alpha$ 
in MCMG with different $N_c$ where $S = 2$.}
\end{figure}

Moreover, the variance tends to a constant value when the control parameter 
$\alpha \rightarrow \infty$. But is such a constant value consistent 
with the coin-toss case value as in MG? Table 1 shows the value of 
$\sigma^{2}$ at $\alpha \approx 100$ in contrast with the coin-toss case value 
for MCMG with different $N_c$.

\begin{table}[h]
\caption{The value of the variance $\sigma^{2}$ at $\alpha \approx 100$ 
and the coin-toss limit value in MCMG with different $N_c$ where $S = 2$.}
\begin{ruledtabular}
\begin{tabular}{ccc} 
$N_c$&$\sigma^{2}$ at $\alpha \approx 100$&coin-toss limit\\ \hline
3 & 0.217 $\pm$ 0.010 & 0.2222 \\ \hline
4 & 0.184 $\pm$ 0.008 & 0.1875 \\ \hline
5 & 0.156 $\pm$ 0.007 & 0.1600 \\ \hline
7 & 0.120 $\pm$ 0.006 & 0.1224 \\ 
\end{tabular}
\end{ruledtabular}
\end{table}

It was shown clearly that the two values agree with each other. In other words, 
the variance does tend to the coin-toss case value as $\alpha \rightarrow 
\infty$ in MCMG with different $N_c$. As a result, we conclude that the variance 
in MCMG, no matter what $N_c$ is, has very similar properties with 
respect to the control parameter as in MG.

Since the variance shows a second order phase transition, it is instructive to 
investigate if $\alpha_o$, the value of $\alpha$ corresponding to minimum
$\sigma^{2}$, depends on $N_c$ or not. The relationship between the minimum
variance $\sigma^{2}_o$ and $N_c$ is also worth to study. We estimated the value
of $\alpha_o$ and $\sigma^{2}_o$ by polynomial interpolation around the point of
minimum variance. The error of $\alpha_o$ and $\sigma^{2}_o$ was 
estimated to be the difference of the value of $\alpha_o$ and $\sigma^{2}_o$ found 
for different degrees of interpolation.

\begin{table}[h]
\caption{The estimated minimum variance $\sigma^{2}_o$ and corresponding 
$\alpha_o$ in MCMG with different $N_c$ where $S = 2$.}
\begin{ruledtabular}
\begin{tabular}{ccc}
$N_c$&$\alpha_o$&$\sigma^{2}_o/(N/N_c)$\\ \hline
3 & 0.574 $\pm$ 0.007 & 0.274 $\pm$ 0.001 \\ \hline
4 & 0.415 $\pm$ 0.009 & 0.272 $\pm$ 0.002 \\ \hline
5 & 0.331 $\pm$ 0.001 & 0.283 $\pm$ 0.002 \\ \hline
7 & 0.308 $\pm$ 0.013 & 0.318 $\pm$ 0.002 \\ 
\end{tabular}
\end{ruledtabular}
\end{table}

Table 2 summarises the estimated values of $\alpha_o$ and $\sigma^{2}_o$ for 
different $N_c$. The estimated values of $\alpha_o$ and $\sigma^{2}_o$ against
$N_c$ with fixed memory size $M$ are also shown in figure~9. We again
divided the variance by $N/N_c$ in order to have an objective comparision of
the variance for different $N_c$. 

\begin{figure}[h]
\includegraphics[scale = 0.28, bb=490 50 300 600]{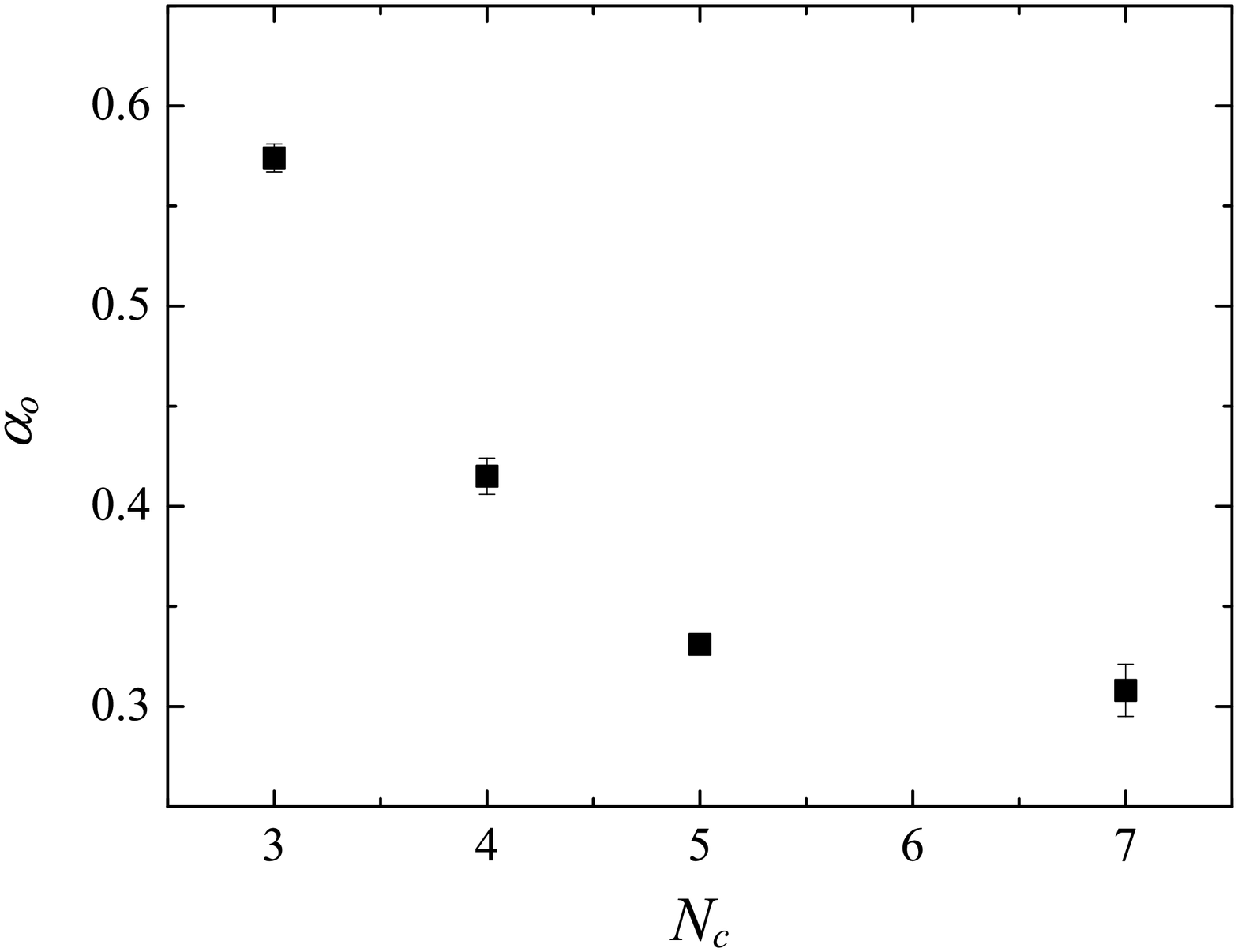}
\end{figure}
\vspace{-8mm}
\begin{figure}[h]
\includegraphics[scale = 0.28, bb=520 50 300 600]{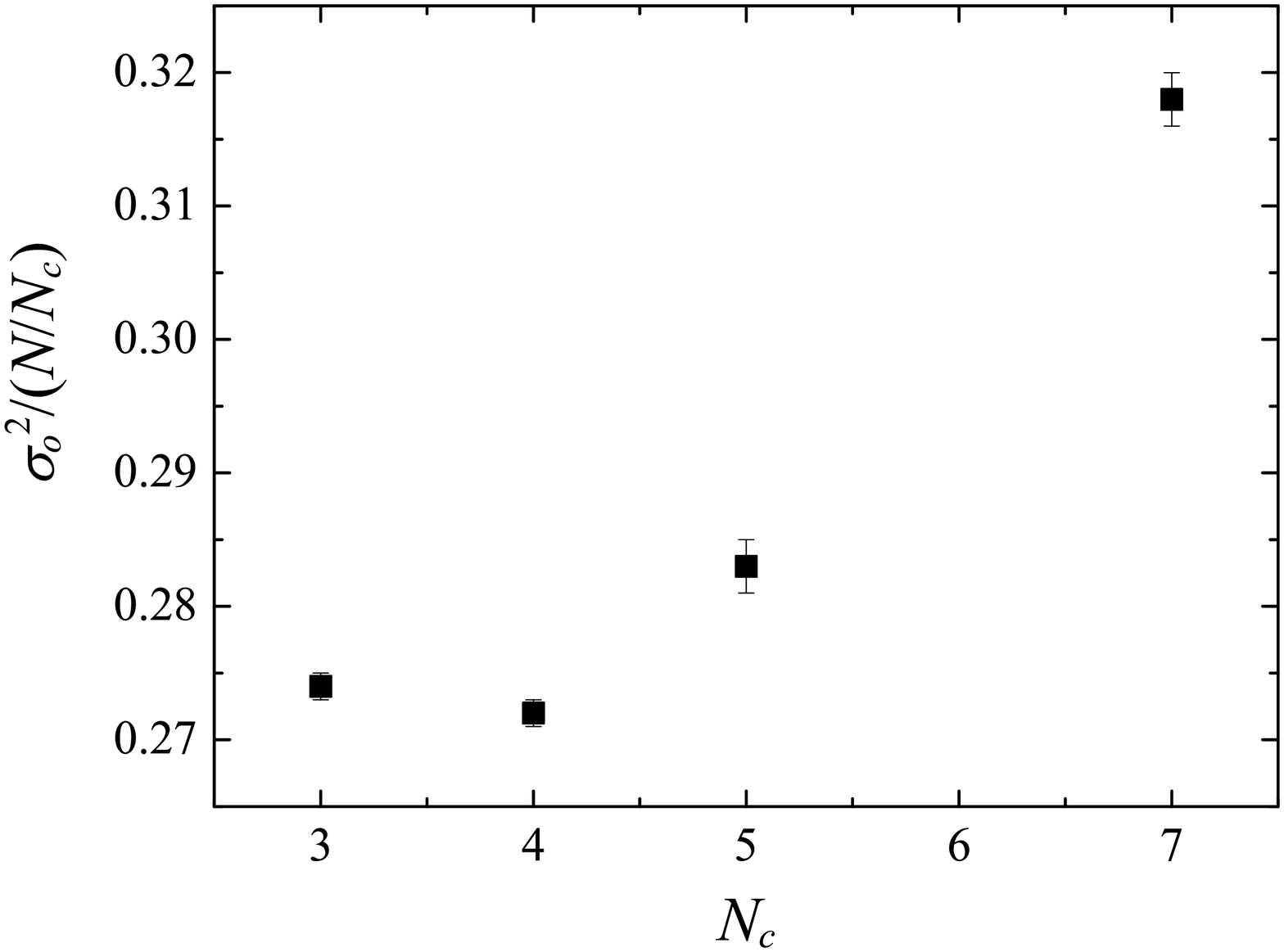}
\caption[1]{Estimated values of $\alpha_o$ and ${\sigma^{2}}_o$ in MCMG with 
different $N_c$ where $S = 2$.}
\end{figure}

We found that in MCMG, $\alpha_o$ decreases as $N_c$ increases. We may explain
this phenomenon as follows:
For MCMG with more number of rooms, more strategies at play are required 
such that players can use all the strategies in the reduced strategy space. Since
maximum cooperation is attained when all the strategies in the reduced strategy
space are used by players, so the value of $\alpha_o$ is larger for MCMG with more
number of rooms.
We also found that the scaled variance $\sigma^{2}_o/(N/N_c)$ increases as $N_c$
increases. It is due to the increase of difficulty for players to cooperate with
each other when there are more choices for them.

On the other hand, we also studied the behavior of the attendance as a function 
of the control parameter $\alpha$ in MCMG with different number of strategies 
$S$. Figures~10 and~11 display the results for $N_c = 3$. The two figures 
indicate that the behavior of the attendance as a function of the control 
parameter $\alpha$ for different number of strategies $S$ are similar, just 
like the case of MG and MCMG with $N_c = 2$. In conclusion, the attendance in 
MCMG has very similar properties with respect to the control parameter as in 
MG no matter how many choices and strategies players have.

\begin{figure}[h]
\includegraphics[scale = 0.28, bb=520 50 300 600]{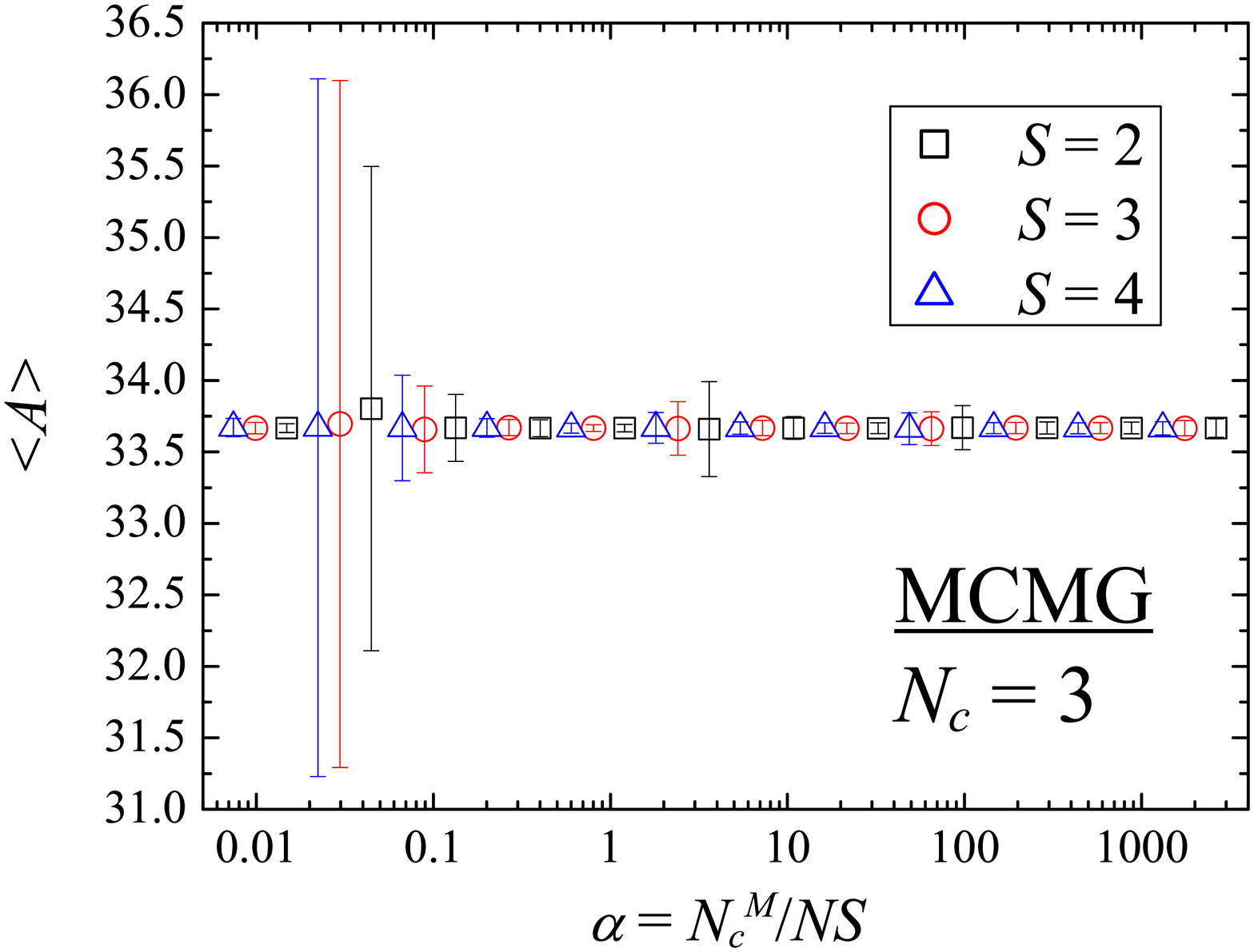}
\caption[1]{The mean attendance $\langle A \rangle$ versus the control parameter 
$\alpha$ in MCMG with $N_c = 3$ where $N = 101$.}
\end{figure}
\begin{figure}[h]
\includegraphics[scale = 0.28, bb=520 50 300 600]{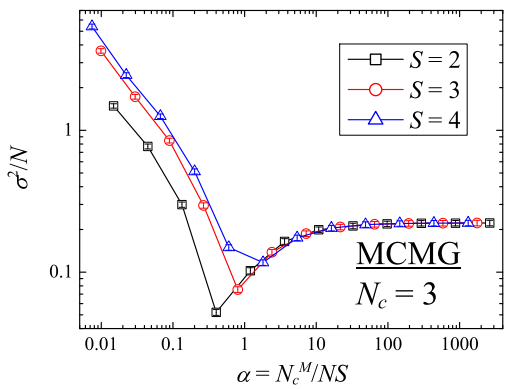}
\caption[1]{The variance $\sigma^2$ versus the control parameter $\alpha$ in MCMG 
with $N_c = 3$ where $N = 101$.}
\end{figure}

\subsection{Comparison of player's wealth for different $N_c$}
We proceed to study the properties of player's wealth in the MCMG. The mean and 
maximum player's wealth as a function of the control parameter $\alpha$ for 
$N_c = 3$ to $7$ were shown in figures~12 and~13.

\begin{figure}[h]
\includegraphics[scale = 0.28, bb=520 50 100 760]{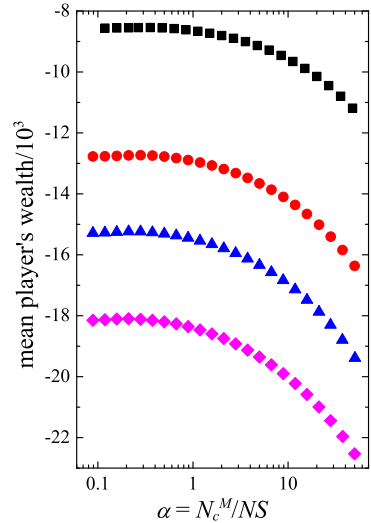}
\caption[1]{Comparison of the mean player's wealth versus the control parameter 
$\alpha$ in MCMG with different $N_c$ where $S = 2$.}
\end{figure}

\begin{figure}[h]
\includegraphics[scale = 0.28, bb=520 50 100 760]{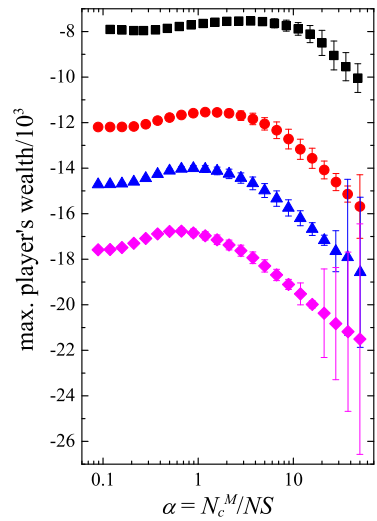}
\caption[1]{Comparison of the maximum player's wealth versus the control
parameter $\alpha$ in MCMG with different $N_c$ where $S = 2$.}

\end{figure}
 
From figures~12 and~13, we notice that player's wealth displays similar 
behavior for MCMG with different number of rooms $N_c$. When the reduced 
strategy space size is much smaller than the number of strategies at play, the 
mean player's wealth remains almost constant as the control parameter $\alpha$
increases. When $\alpha$ is small, the system is in the overcrowding phase
where all the rooms have the same chance to ``win". Therefore, all the players,
on average, always win the same number of times and make the same amount of
profit for small $\alpha$. Then the mean player's wealth attains a maximum
value near the point of minimum variance when the reduced strategy space size
is approximately equal to the number of strategies at play. If the control
parameter $\alpha$ increases further, the number of players $N$ is comparable
to the number of rooms $N_c$. Thus, finite size effect is important. In this
case, some of the rooms will have a higher
chance to ``win". Then some of the players will always lose while some of them
will always win in the game. As a result, the mean player's wealth decreases
rapidly when the control parameter $\alpha$ increases further.

The properties of the maximum player's wealth is similar to the mean player's 
wealth except there is a significant peak correponding to maximum 
cooperation of players. 
Although players on average always perform the same for small $\alpha$, but
the smart players can perform much better when the small control parameter
$\alpha$ increases. Therefore, those smart players is wealthier when the
control parameter $\alpha \rightarrow 1^+$. 

\section{Conclusions}
\label{SecConc}
Although the MCMG and MG models are not exactly the same, our work shows that 
the attendance of a room in MCMG has similar behavior to that in MG as a 
function of the control parameter. Besides, we found that the 
attendance of different rooms displays almost the same behavior in MCMG with 
number of choices $N_c \ge 2$. Moreover, we observed that both the 
attendance and player's wealth displays similar properties as a function of the 
control parameter in MCMG with different $N_c$. As all the above 
mentioned features can be explained reasonably, so we concluded that we have 
successfully built a computationally feasible model of multi-choice minority 
game --- MCMG. 

Various extensions of the MCMG model could be studied. For example, the 
multi-choice game with zero-sum and the multi-room game with agents who can 
invest different amount. We hope the study of the extensions of the MCMG model 
can give us more insight on more realistic complex adaptive system.

\begin{acknowledgments}
We would like to thank P.~M. Hui and Kuen Lee for their useful discussions and 
comments, especially during the annual conference of the Physical Society of 
Hong Kong in June 2001. F.~K.~C. would also like to thank K.~M. Lee for his 
useful discussions and comments. 
This work is supported by in part by the RGC grant of the Hong Kong SAR
government under the contract number HKU7098/00P. H.~F.~C. is also supported by
the Outstanding Young Researcher Award of the University of Hong Kong.
\end{acknowledgments}

\end{document}